\documentclass{PoS}

\title{Prospects for beyond Standard Model Higgs boson searches at future LHC runs and other machines}

\ShortTitle{Prospects for BSM Higgs searches at future LHC and other machines}

\author{\speaker{Krisztian Peters}%
       \\
      Deutsches Elektronen-Synchrotron (DESY)\\
      E-mail: \email{krisztian.peters@desy.de}}

\abstract{The discovery of the Higgs boson at the LHC opened up a new possibility to obtain a precise understanding of the origin of mass and the structure of the vacuum. Several theories beyond the Standard Model predict an extended Higgs sector, with additional scalar particles. We outline the complementary search possibilities for these beyond Standard Model Higgs bosons at the LHC and at possible future high-energy collider facilities.}

\FullConference{Prospects for Charged Higgs Discovery at Colliders\\
		 3-6 October 2016\\
		 Uppsala, Sweden}

\begin{document}

\section{Introduction}

The discovery of the Higgs boson at the LHC is a huge leap forward in understanding electroweak symmetry breaking and the fundamental theory of particle physics. Although within the current experimental precision the discovery confirms and completes the Standard Model (SM) of electroweak and strong interactions, it also raises its shortcomings as the predicted Higgs boson mass becomes divergent due to radiative corrections. Finding the correct extension of the SM which cures its high-energy behaviour is one of the current major fundamental goals of particle physics. Furthermore, despite the advancements with the Higgs boson discovery, the origin of electroweak symmetry breaking is not yet fully answered. A precise understanding of the Higgs sector and finding evidence for physics beyond the SM (BSM) are intimately related and searching for further scalar particles is an important task at the LHC and future high-energy particle colliders. 

Several experimental signatures can be explored in the search for an extended Higgs sector. First of all, the most important question to answer is if the discovered Higgs boson with a mass of around 125 GeV (denoted as $h(125)$) is fully responsible for electroweak symmetry breaking or if it is part of a larger extension of the SM Higgs sector. These questions can be explored with the precise measurements of the $h(125)$ properties. These property measurements will be complemented with direct searches for additional Higgs bosons, such as predicted for example in two-Higgs doublet models. Finally, one can also use the newly discovered boson to search for particles or interactions not predicted by the SM. These can be particles whose decay chain leads to $h(125)$ in the final state, such as new high-mass resonances decaying to Higgs boson pairs or flavour changing neutral currents with decays to $h(125)$. Further non-SM Higgs boson decays can be for example lepton flavour violating decays or Higgs portal models. In these models, dark matter particles interact with SM particles via Higgs boson exchange.

The current main collider physics experiments to explore all these questions are the general purpose experiments ATLAS and CMS at the Large Hadron Collider (LHC). At the moment, with the LHC Run-2, proton-proton collisions are ongoing with a centre-of-mass energy of 13 TeV and the LHC is going to run for another two decades with exciting physics prospects. Especially for precision measurements of $h(125)$, a future electron-positron collider would be of great advantage and several possible projects are under discussion. These are two linear collider projects, the International Linear Collider (ILC) and the Compact Linear Collider (CLIC). And two circular colliders, the Future Circular Collider (FCC-ee) and the Circular Electron Positron Collider (CEPC). These two circular colliders would serve as an intermediate step to reach for significantly higher energies with proton-proton colliders in the same tunnels following the lepton machines, named FCC-hh and Super Proton-Proton Collider (SppC), respectively. In the following article we give an overview of these different collider prospects and discuss examples for possible BSM Higgs boson search reaches.

\section{The High Luminosity LHC}

The aim of the LHC and the two general purpose detectors is to accumulate a data sample of over 3 ab$^{-1}$ by the end of 2037. The current LHC Run-2 will end in 2018 and during a two year long shutdown the injector, the LHC and the experiments will realise their Phase-1 upgrade in order to reach the design instantaneous luminosity of ${\cal L} = 2 \cdot 10^{34} cm^{-2}s^{-1}$. Run-3 will start in 2021, and the accelerator will also reach the design collision energy of 14 TeV. In 2024 the LHC and experiments will have the Phase-2 upgrade to prepare for the High Luminosity LHC (HL-LHC). With an instantaneous luminosity of ${\cal L} = 5 - 7.5 \cdot 10^{34} cm^{-2}s^{-1}$ the LHC will deliver the entire Run-2/3 dataset per year, to reach a final integrated luminosity of over 3~ab$^{-1}$. 

\begin{figure}[t]
\centering
\includegraphics[width=0.45\linewidth]{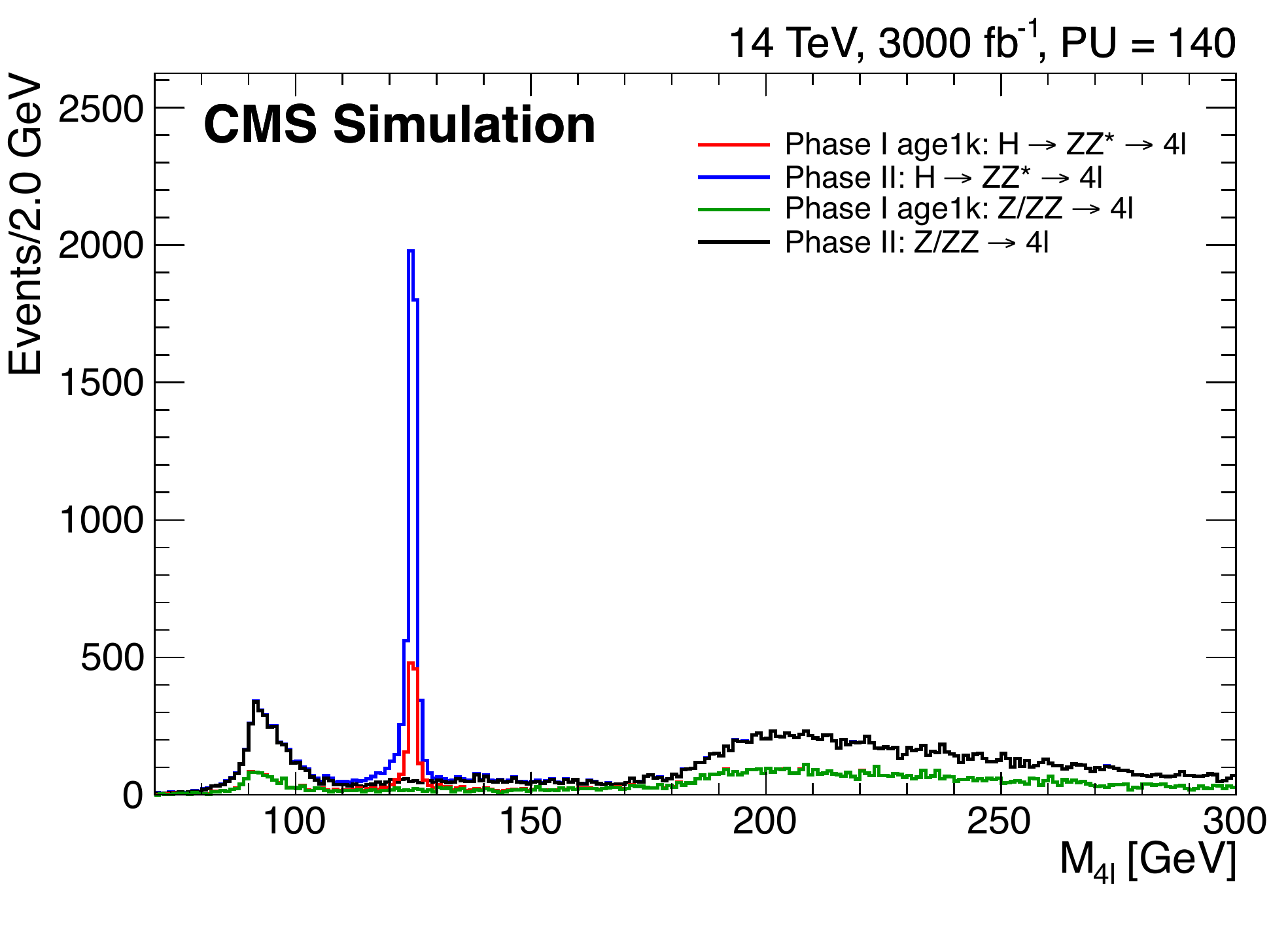} (a)
\includegraphics[width=0.45\linewidth]{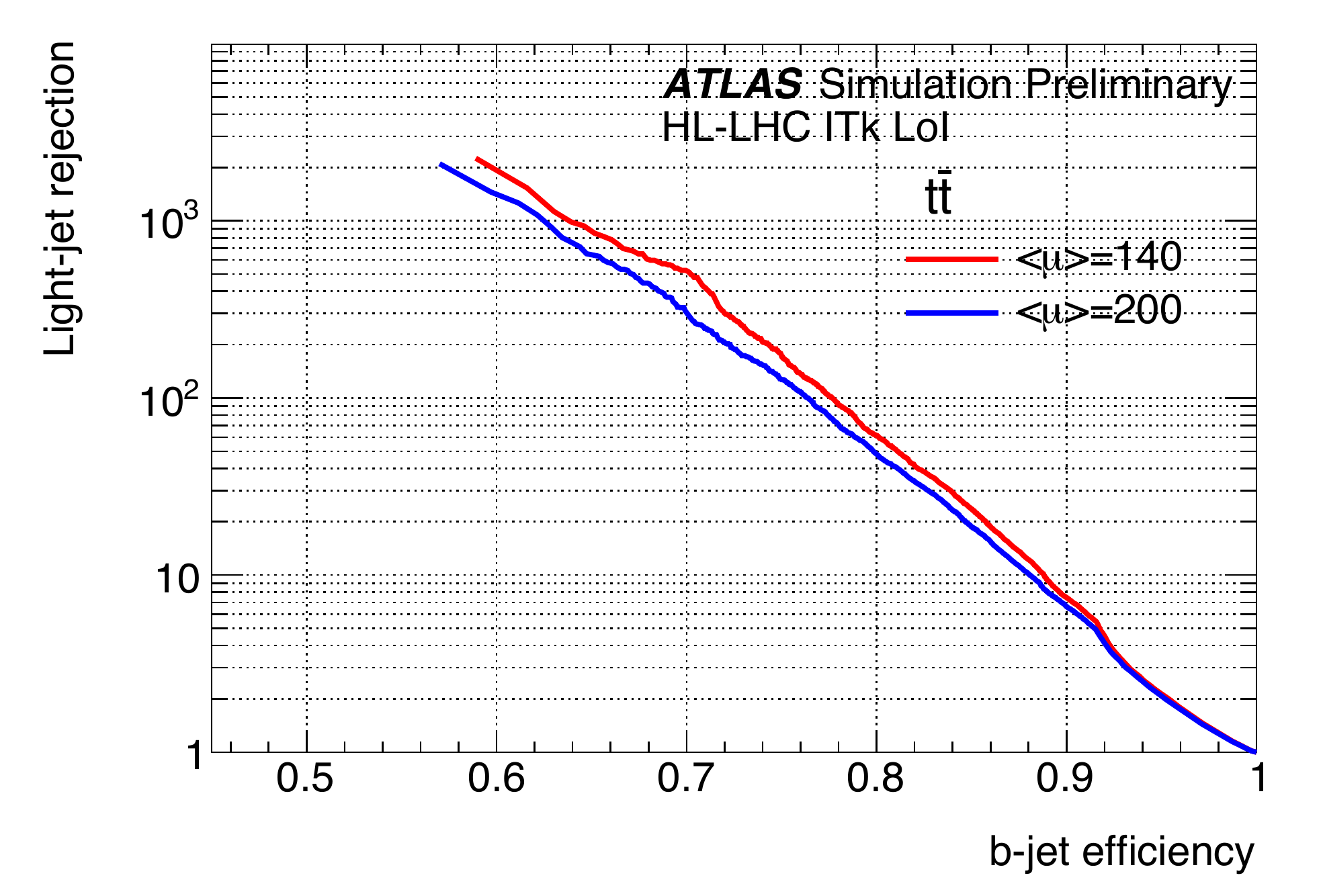} (b)
\caption{(a) Four lepton mass distributions at CMS for the full Phase-II luminosity of 3000 fb$^{-1}$
for the Higgs boson signal and background. Both processes have been simulated with the aged Phase-I detector with pileup of 140 and the Phase-II detector with pileup of 140, \cite{CMSupgrade}. (b) The light-flavour jet rejection vs the b-tagging efficiency at the ATLAS Phase-II detector with pileup of 140 and 200 for jets with $p_T > 20$ GeV and $|\eta|<2.7$, \cite{ATLASupgrade}. }
\label{fig:upgrade}
\end{figure}

ATLAS and CMS are well understood  detectors, they have a stable operation and data taking efficiencies above 90\%. Nevertheless, for the future LHC runs several upgrades are mandatory. Both experiments were designed for an average number of 20 - 30 proton-proton collisions per bunch crossing (pileup). At the HL-LHC it is expected that the average pileup will rise to 140 - 200. Major upgrades are nessesary for the experiments in order to cope with these challenging experimental conditions and the radiation damage caused over the time, yet maintain similar levels of performance as of today. Figure \ref{fig:upgrade}(a) illustrates the Higgs boson to four lepton decay signal, which compares the expectations from an upgraded CMS detector to the one which would be obtained with the current detector, aged by the expected total radiation damage. 

The upgrade plans are similar for ATLAS and CMS. These involve a complete replacement of the inner detector with a fully silicon-based tracker. The exact layout is currently under study, however both experiments consider an extended coverage to higher rapidities, which helps the pileup mitigation in the forward region and is also of advantage for an important part of the physics programme (e.g. vector boson scattering). The new tracking detectors will have a higher granularity in order to reduce the occupancy with several multiple collisions and they will have a better radiation tolerance compared to the current detectors. New trigger and data-acquisition systems are planed to enable trigger rates up to an order of magnitude larger than with the current systems. This requires replacing almost all of the current readout electronics. Furthermore, there will be significant upgrades to the muon system and to calorimetry in the forward region. Figure \ref{fig:upgrade}(b) shows an example of a performance evaluation of the upgraded ATLAS tracking system based on full simulation. The plot shows the expected b-jet efficiency versus the light-jet rejection for two pileup scenarios. One can note a similar performance compared to Run-2 for a pileup of  $\mu = 200$, and a 30\% better rejection for $\mu = 140$. For this evaluation the current reconstruction software is used, hence it is expected that these performances will further improve in the future. Since no appreciable performance losses are expected despite the challenging environment, current search reaches and measurement precisions can be extrapolated to a larger dataset for a first rough assessment. 

\begin{figure}[t]
\centering
\includegraphics[width=0.9\linewidth]{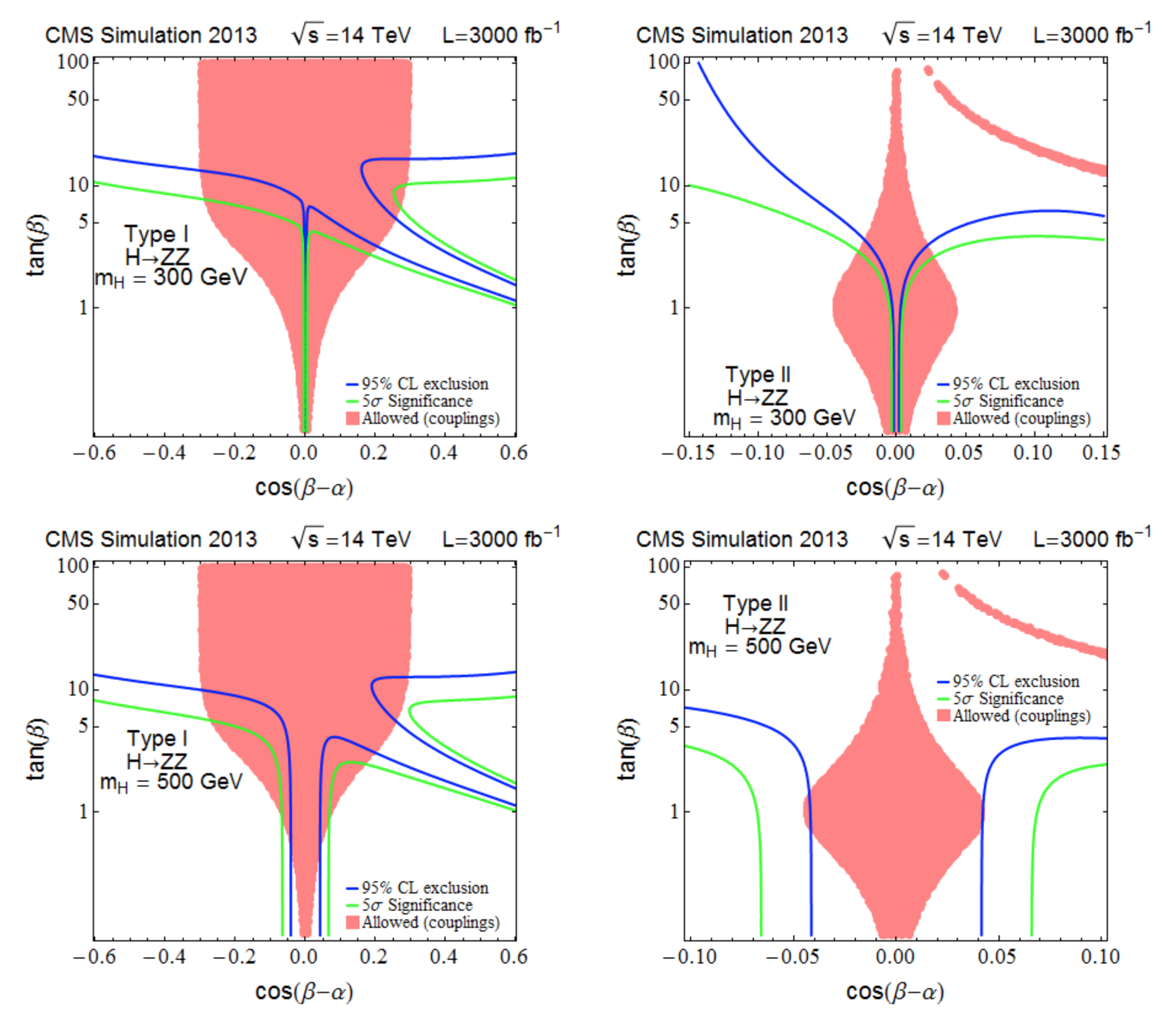} 
\caption{The region of parameter space for which a 300 GeV (top) and 500 GeV (bottom) H boson could be excluded at 95\% CL (blue), and the region of parameter space which could yield a 5$\sigma$ observation of a heavy scalar H boson (green) in the ZZ channel, in the context of Type I (left) and II (right) 2HDMs \cite{2hdmexclusion}. The coloured regions correspond to the expected 95\% CL allowed region from Higgs precision measurements with 3000 fb$^{-1}$ of data \cite{heather}.}
\label{fig:2hdm}
\end{figure}

The main objectives of Higgs boson physics at the HL-LHC is to measure the couplings of $h(125)$ with a few percent accuracy; to gain access to Higgs pair production and rare decays; and to search directly for heavy Higgs partners. At the HL-LHC, Higgs couplings will be measured down to $\sim$5\% accuracy, assuming a reduction of the current theory uncertainties. 
A prominent example for rare decays is the measurement of the muon-Yukawa coupling, which will reach only meaningful precision at the HL-LHC. It will be the first time the Yukawa coupling of a second generation fermion will be scrutinized. 

\begin{figure}[t]
\centering
\includegraphics[width=0.55\linewidth]{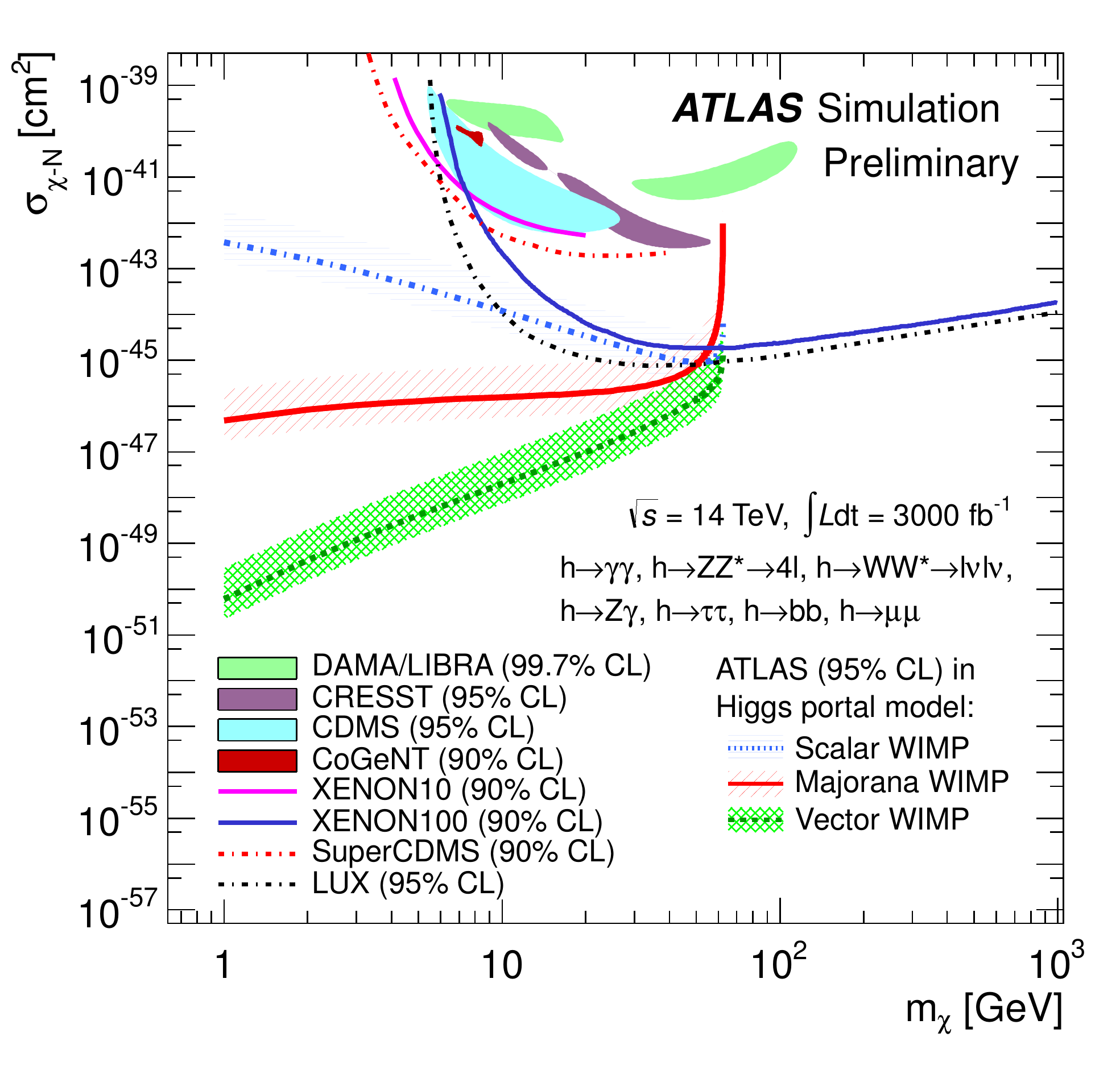}
\caption{ATLAS expected upper limit at 95\% CL on the WIMP-nucleon scattering cross section in a Higgs portal model as a function of the mass of the dark matter particle, shown separately for a scalar, Majorana fermion, or vector boson WIMP, with 3000 fb$^{-1}$ of data and including all systematic uncertainties, \cite{portal}. The hashed bands indicate the uncertainty resulting from the form factor fN. Excluded and allowed regions from direct detection experiments at the confidence levels indicated are also shown.  These are spin-independent results obtained directly from searches for nuclei recoils from elastic scattering of WIMPs, rather than being inferred indirectly through Higgs boson exchange in the Higgs portal model.}
\label{fig:higgsportal}
\end{figure}

The $h(125)$ precision measurements and searches for additional Higgs bosons are highly complementary as it can be illustrated in the example of a general two-Higgs double model (2HDM). In these models an additional Higgs doublet is introduced, yielding five Higgs bosons (two of which are charged). Such a Higgs sector is described by the Higgs boson masses, the ratio $\tan \beta$ of the vacuum expectation values of the two doublets and the mixing angle $\alpha$ between the two CP-even states. At tree level the two free parameters $\alpha$ and $\beta$ can be directly related to the coupling scale factors of $h(125)$, hence the precise measurements of $h(125)$ constrain the 2HDM parameters $\alpha$ and $\beta$, independently of the masses of the heavier Higgs bosons. Sensitivity extrapolations to such constraints are shown in Fig. \ref{fig:2hdm} for Type I and Type II models. The red area shows the expected parameter space to be excluded by the end of the HL-LHC. In the alignment limit, where $\cos (\beta - \alpha)$ approaches zero, the $h(125)$ boson has exactly the SM Higgs couplings. This region of the parameter space is naturally the most difficult one to exclude. Some of this parameter space will be more accessible with direct searches for additional Higgs bosons as can be seen in the same plot. The lines show the exclusion ranges for direct searches for the CP-even Higgs boson $H$ in gluon-fusion production in the $ZZ$ decay mode. This analysis is powerful in the low $m_H$ and $\tan \beta$ region, otherwise the top-anti-top branching ratio dominates and the $bbH$ production mode is also important. The analysis assumes that higher Higgs boson masses are degenerate and that the width of the $H$ boson is smaller that the experimental resolution.



A further example for searches of a BSM Higgs sector is the Higgs portal model, which again illustrates nicely the interplay between precision measurements and direct searches. Upper limits on the Higgs boson branching ratio to weakly interacting particles (called invisible branching ratio) can be derived using the combination of rate measurements of all visible decays and assume SM couplings to visible particles. With this the interpolation to the HL-LHC constraints give an invisible branching ratio limit of $BR_i < 0.13 (0.09)$ at 95\% CL (without) theory uncertainty for 3 ab$^{-1}$ at ATLAS. This result can be converted to (model dependent) limits on dark matter interactions, such as shown in Fig. \ref{fig:higgsportal}, which is nicely complementary to direct dark matter search experiments. On the other hand, one can search directly for the invisible decay signature. The most sensitive search channel for direct detection is the vector boson fusion signature. Exclusion limits are expected to reach $BR_i < 0.05$ at 95\% CL, however the  improvement of theory uncertainties will be crucial for this search. The ultimate sensitivity will come from future $e^+e^-$ colliders. 

\begin{figure}[t]
\centering
\includegraphics[width=0.8\linewidth]{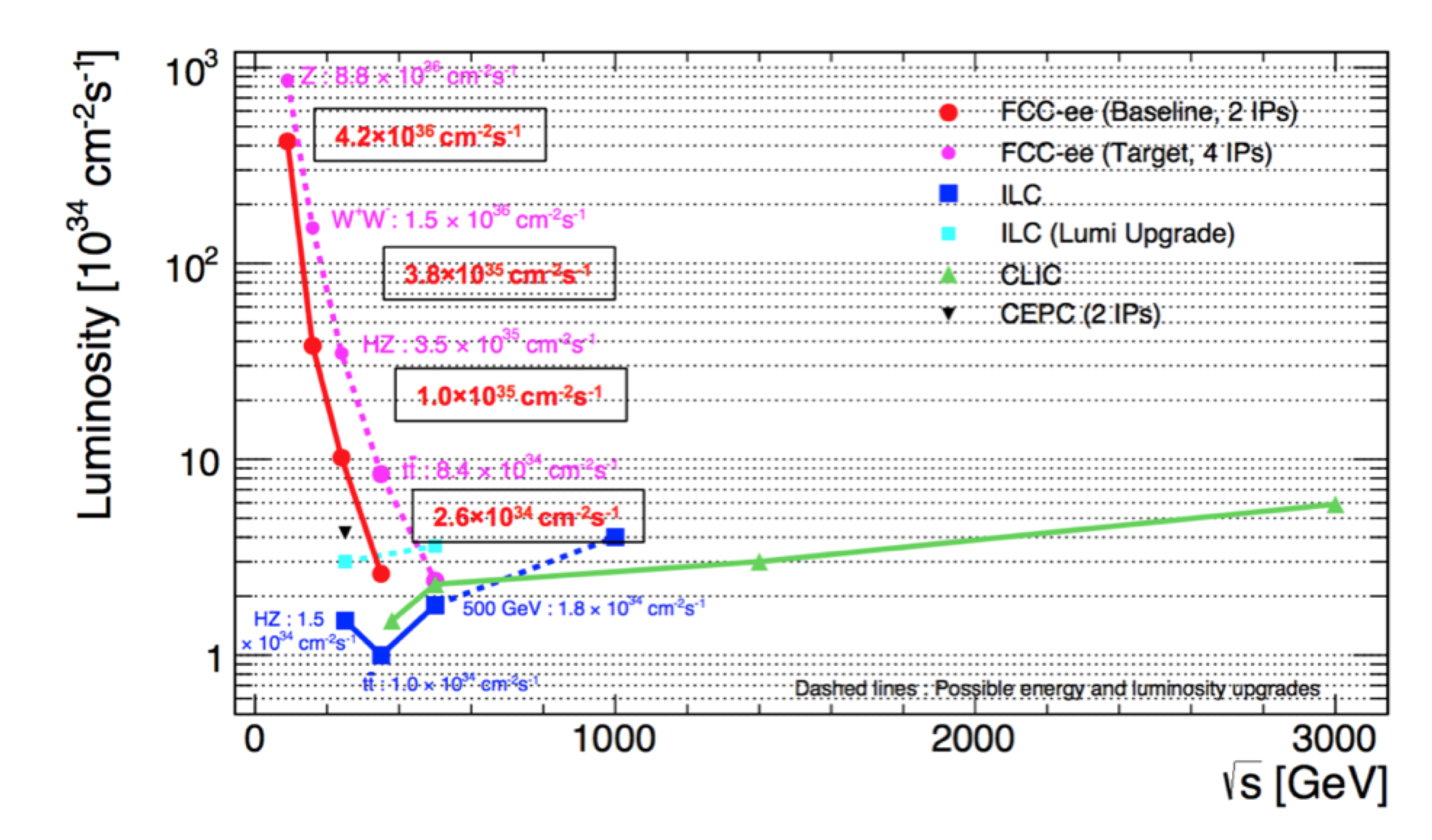}
\caption{Target luminosities as a function of center-of-mass energy, for circular (FCC-ee, CEPC) and linear (ILC, CLIC) $e^+e^-$ colliders currently under consideration.}
\label{fig:lumi}
\end{figure}

\section{Future colliders}

The ILC is technically the most advanced future $e^+e^-$ collider project. Its technical design report was published in 2013 and the site in Japan is chosen. Currently the project is reviewed by the Japanese science ministry (MEXT). The ILC can be operated with (partially) polarised beams, with a center-of-mass energy of up to 500 GeV, which can be upgraded to 1 TeV. The 500 GeV version would have a total length of 34 km. With an intermediate luminosity upgrade, the 20 year long running programme foresees an integrated luminosity of 2 ab$^{-1}$ with 250 GeV collisions and 4 ab$^{-1}$ at 500 GeV. 

Higher collision energies of up to 3 TeV would be achieved with CLIC, which has a proposed site near Geneva. The physics prospects are comparable to the ILC, however the higher centre-of-mass energies would increase the BSM reach in direct searches. A conceptual design report (CDR) was published in 2012. Some of the main challenges arise from the nm-size beams and the power consumption at 3 TeV, R\&D in several areas is still ongoing.

The FCC project has also a proposed site near Geneva at CERN. The main goal is a 100 TeV $pp$ machine (FCC-hh), which would deliver in 10 years  roughly the HL-LHC total luminosity, and an ultimate integrated luminosity of 20 ab$^{-1}$ after 25 years of operation. This collider would have roughly a 100 km circumference with 16 Tesla high-field superconducting magnets, a layout which would also fit the geology of the area. An $e^+e^-$ collider (FCC-ee) can be installed as an intermediate step, with collision energies tunable from the $Z$ boson mass to $O(365)$ GeV. The lepton collider would have a top-up injection scheme from a separate storage ring, similar to the B-factories. For the FCC project, the CDR is expected by the end of 2018, in time for the next European Strategy Update in 2019. While the hadron collider would open up a new high-energy regime for discoveries, the lepton collider would be an excellent option for precision studies due to the very high integrated luminosities and low beam energy spread which are possible at a circular collider. It is estimated that with a conservative optics and two interaction regions the FCC-ee would produce in 3 years $5\cdot 10^{12}$ $Z$ bosons and in 5 years 1 million Higgs bosons. 

The CEPC, followed by the SppC, is a similar proposal for a circular collider project in China. Currently a collider circumference of 54 km is proposed, which would allow the $e^+e^-$ collider to reach energies up to 250 GeV and the $pp$ collider to reach energies of 70 TeV. In contrast to FCC-ee, the proposed $e^+e^-$ project would constitute a one-ring machine to accelerate electrons and positrons, which would collide head-on (i.e. without a crossing angle), leading to a more challenging design.

Figure \ref{fig:lumi} compares the energy and luminosity reach of the different $e^+e^-$ collider options. In general, circular colliders have the advantage of higher luminosities, while linear colliders can reach higher energies. Besides these two parameters there are many other relevant aspects, such as the amount of beamstrahlung and beam energy spread, the possibility of having polarised beams, the bunch structure (which is important for the detectors) and the number of possible interaction points. 

\begin{figure}[t]
\centering
\includegraphics[width=0.6\linewidth]{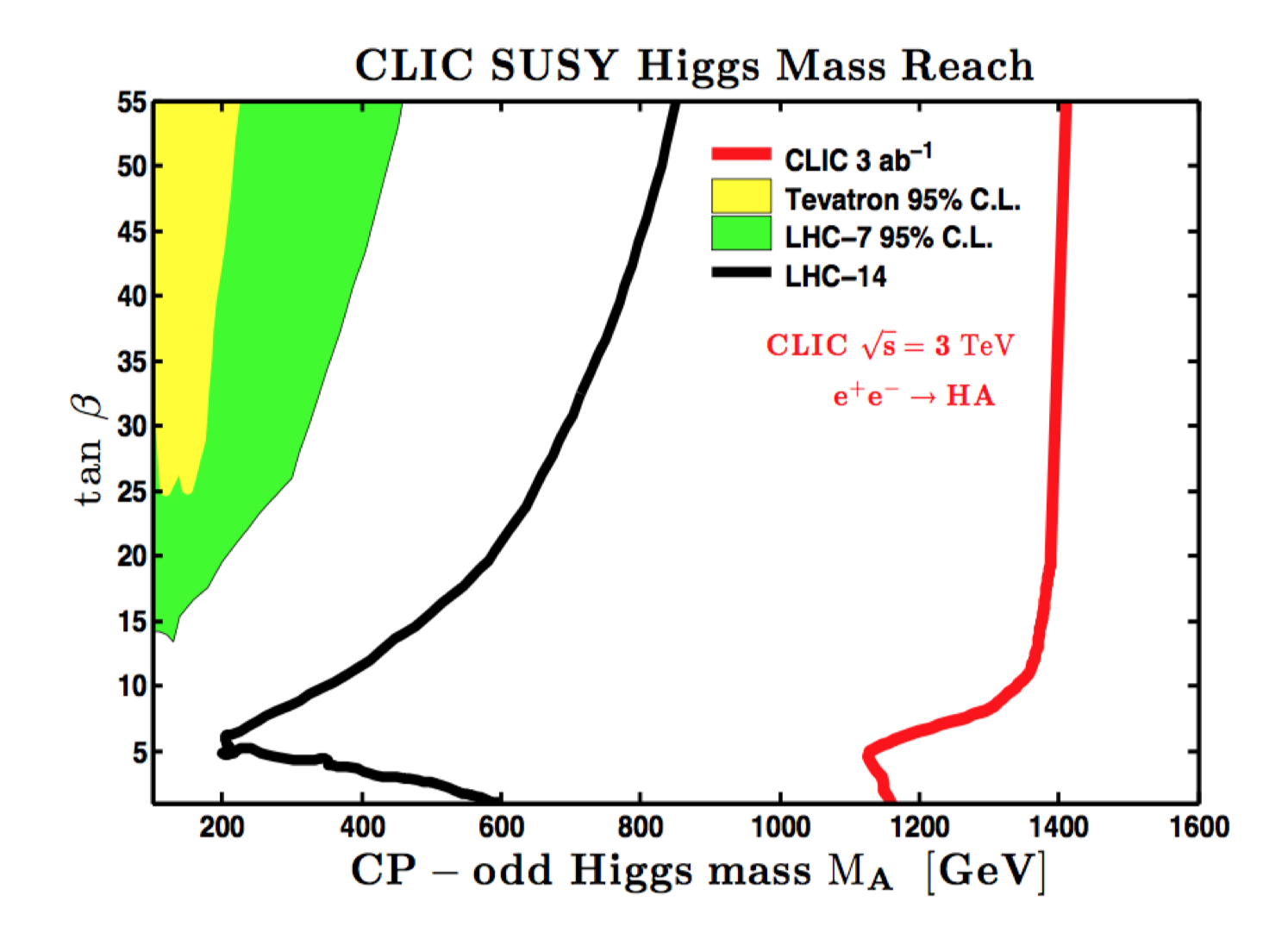}
\caption{Projected exclusion regions for a CP-odd Higgs boson $A$. The black line is the projection of the search reach at the LHC with $\sqrt s = 14$ TeV and 300 fb$^{-1}$ of luminosity. The red line is the search reach of CLIC with $\sqrt s = 3$ TeV \cite{clic}. A linear collider at $\sqrt s = 500$ GeV can find heavy Higgs mass eigenstates if their masses are below the kinematic threshold of 250 GeV.}
\label{fig:clic}
\end{figure}

For Higgs production,  lepton colliders have the advantage of low backgrounds, hence all decay modes are accessible (e.g. $h\to cc$). Coupling measurements can be done in a model independent fashion, as the total decay width is measurable. The study of associated production with top quarks ($t\bar th$) and di-Higgs production ($hh$) in $e^+e^-$ collisions require $\sqrt s > 500$ GeV, which is only possible at the linear colliders. Future hadron colliders will have the advantage of huge cross-sections and a high-energy reach, which are advantageous for clean/rare decays and heavy final states. Due to large backgrounds however, not every decay will be accessible. For coupling measurements within the kinematic reach the best precision will be obtained at the FCC-ee due to high luminosity and low beam energy spread. For heavy states ($t\bar th$ and $hh$) higher energies are needed (ILC/CLIC, FCC-hh/SppC). Especially between the future lepton and hadron colliders there is a very nice complementarity in Higgs physics. While in general lepton colliders will reach higher coupling precisions, better measurements will be achieved at hadron colliders for the top-Yukawa coupling and the Higgs boson self-coupling, as well as for clean and rare decays, such as $h \to \gamma \gamma$ and $h \to \mu \mu$. 

\begin{figure}[t]
\centering
\includegraphics[width=0.88\linewidth]{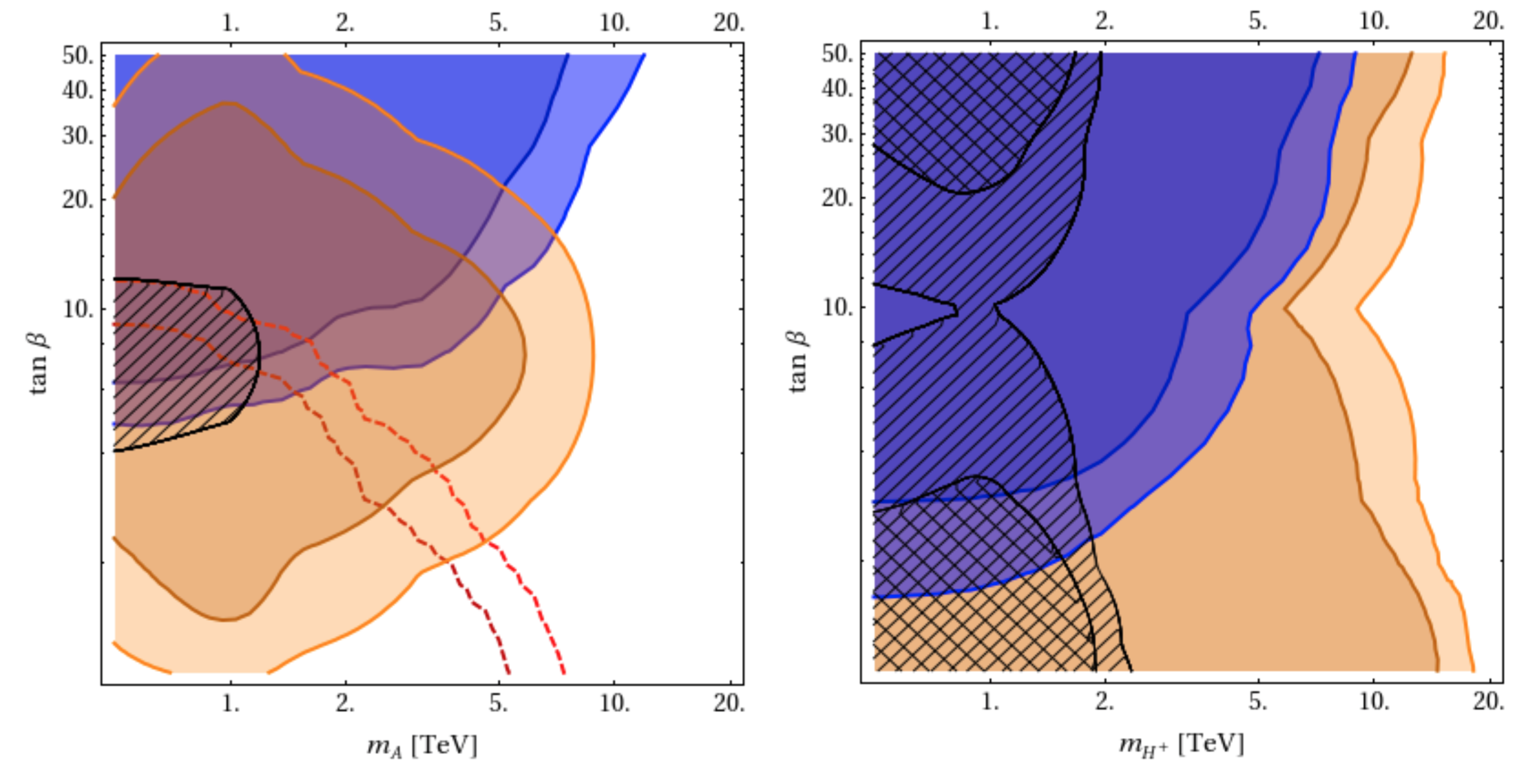}
\caption{Exclusion limits for MSSM Higgs bosons. For the CP-odd neutral Higgs boson, the blue and orange regions are probed by the channels $pp \to bbA \to bb\tau \tau$, $pp \to bbA \to bbtt$, respectively. For the charged Higgs bosons, the blue and orange regions are probed by the channels $pp \to tbH^\pm \to tb\tau \nu$ and $pp \to tbH^\pm \to tbtb$, respectively. Given the same channel or the same color, the two different opacities indicate the sensitivities w.r.t. a luminosity of 3 ab$^{-1}$ and 30 ab$^{-1}$ at a 100 TeV $pp$ collider, respectively. The cross-hatched and diagonally hatched regions are the predicted exclusion contours for associated Higgs production at the LHC for 0.3 ab$^{-1}$, and 3 ab$^{-1}$, respectively, \cite{Hajer:2015gka}.}
\label{fig:fcc}
\end{figure}

This precision which will be achievable on $h(125)$ coupling measurements at lepton colliders will provide a powerful probe for heavy Higgs bosons as well. This was for example studied for the pMSSM case in Ref. \cite{Cahill-Rowley:2014wba}, where it is shown that precision measurements at a lepton collider can strongly constrain the existence of heavier Higgs bosons up to a few TeV.  Direct searches for heavier Higgs bosons are somewhat limited by the lower collider energies. Since the main production mode for charged and neutral ($A$, $H$) Higgs bosons is pair-production, the mass reach is roughly constrained to half of the centre-of-mass energy. This can be seen by the exclusion limit plot shown in Fig. \ref{fig:clic}, where masses up to $\sim 1400$ GeV can be excluded by operating CLIC at the highest energies of 3 TeV. If a new particle is within the production reach of lepton colliders, its properties can be studied very precisely due to the low backgrounds and the simpler collision environment. 

At a future hadron collider, e.g. at the FCC-hh, the reach for direct searches for heavy Higgs bosons get strongly extended compared to the LHC simply from the large cross section enhancement. This enhancement is a factor of $\sim$30 - 50 for $m_H = 500$ GeV, depending on the exact parameters of the model. The production of charged Higgs bosons gets enhanced a factor $\sim$90 accordingly. An estimate of future exclusions ranges for the LHC and FCC-hh are compared in Fig. \ref{fig:fcc} for a CP-odd Higgs and a charged Higgs boson in the MSSM.

\section{Conclusions}

A precise understanding of the structure of the vacuum and of the origin of electroweak symmetry breaking are among the main goals in particle physics. The LHC experiments are entering the area of precision measurements and rare phenomena searches. 
With the HL-LHC and a significantly increased dataset we will see many more results from the LHC in the next two decades. The ultimate precision on most of the properties of $h(125)$ will be delivered by  an $e^+e^-$ collider and there are several excellent options on the market. A future hadron collider with energies up to the 100 TeV range will significantly enhance the direct discovery reach for BSM Higgs bosons. The study of a BSM Higgs sector will be in many ways complementary, between direct searches and $h(125)$ precision measurements and between hadron and lepton colliders, which provides exciting future scientific possibilities.

\end{document}